# A Scheme for Deterministic *N*-photon State Generation Using Lithium Niobate on Insulator Device

Hua-Ying Liu[a, †], Minghao Shang[a, †], Xiaoyi Liu[a], Ying Wei[a], Minghao Mi[a], Lijian Zhang[a], Yan-Xiao Gong[a, *], Zhenda Xie[a, *], And Shi-Ning Zhu[a]

[a]National Laboratory of Solid State Microstructures, School of Electronic Science and Engineering, School of Physics, College of Engineering and Applied Sciences, and Collaborative Innovation Center of Advanced Microstructures, Nanjing University, Nanjing 210093, China

**Abstract**. Large-photon-number quantum state is a fundamental but non-resolved request for practical quantum information applications. Here we propose an *N*-photon state generation scheme that is feasible and scalable, using lithium niobate on insulator circuits. Such scheme is based on the integration of a common building block called photon-number doubling unit (PDU), for deterministic single-photon parametric down-conversion and up-conversion. The PDU relies on $10^7$-optical-quality-factor resonator and mW-level on-chip power, which is within the current fabrication and experiment limits. *N*-photon state generation schemes, with cluster and GHZ state as examples, are shown for different quantum tasks.

*Address all correspondence to Yan-Xiao Gong, E-mail: gongyanxiao@nju.edu.cn; Zhenda Xie, E-mail: xiezhenda@nju.edu.cn
†These authors contributed equally to this work.

Quantum information has the potential to bring revolutionary performance for the computation, communication, and metrology applications, in terms of speed,[1, 2] security,[3, 4] and accuracy.[5] Similar to the case of classical information, where the system capability relies on the large number of classical bits, the number of qubits is the key resource that defines the overall performance of a quantum system. Though the quantum source benefits from the superposition of an *N*-qubit state, and boosts the information exponentially over an *N*-bit classical source.[6] In fact, a 300-qubit quantum state has more variations than all the atoms in the known universe classically.[7, 8] However, generation of large-size quantum states is a fundamental challenge, no matter for the electrical or optical system. Although the strong interacting particles like superconductivity electrons and ions make it easier to generate large-size quantum states and keep the current record of the qubit number,[9-11] their strong interaction also leads to the decoherence problem and can only work in ultra-low temperature and vacuum environment. These electrical quantum sources still suffer from short state lifetime even under these conditions.

On the other hand, the photons are known for their weak interaction, where long coherence time can be achieved even at room temperature, which makes them suitable for "flying qubits" applications.[12-15] However, they have not been considered as a good candidate to build a large-size quantum source due to their weak interaction in the normal medium. Nonlinear optical medium is by far the most effective way to establish interaction between photons. Using spontaneous parametric down-conversion (SPDC)[16] or spontaneous four-wave mixing (SFWM),[17] two-photon states can be generated directly, but only probabilistically. Even though, it is still an interesting topic for the generation of $N$-photon states. Much effort has been devoted, for their non-deterministic generation, like post-selection with multiple SPDC from the same pump source[18] or cascaded SPDC,[19, 20] though the generation rate decreases exponentially as the photon number increases in these studies. An ultimate solution is realizing deterministic two-photon state generation, and then a deterministic $N$-photon state can be achieved by cascading the two-photon processes. Such concept has been proposed and theoretically studied, however, only with the ideal $\chi^{(2)}$ [21] or $\chi^{(3)}$ [22] material assumption.

Here, we propose the first feasible scheme to deterministically generate $N$-photon state, considering the practical material capability. Such scheme is based on an ensemble of basic units called photon-number doubling units (PDUs), which is used to realize photon number doubling and keep their spectrum unchanged at the same time. This unit is capable for deterministic parametric down-conversion (DPDC) and deterministic parametric up-conversion (DPUC). Taking advantage of the strong nonlinear interaction in lithium niobate on insulator (LNOI) circuits,[23-27] this PDU only requires LNOI resonator with ~$10^7$ optical quality factor ($Q$ factor) and mW-level on-chip power, which have already been demonstrated in experiments.[28, 29] With miniaturized footprint in LNOI, PDUs can be integrated for unlimited photon number in principle on a single chip. Therefore, our scheme may fulfill the fundamental demand for scalable large-size quantum state generation, as examples discussed here for cluster and GHZ state generation.

In general, $N$-photon state can be generated by cascading two-photon generation processes, either via $\chi^{(2)}$ or $\chi^{(3)}$ nonlinearity.[22] Here we choose the LNOI circuit to realize $\chi^{(2)}$ parametric down-conversion (PDC) with unitary efficiency, because of its high nonlinearity [29] and strong mode confinement.[30] As shown in Fig. 1, the photon number can be scaled using a standard PDU. The photon number doubling can be achieved using DPDC, while it is also necessary to include a DPUC unit for frequency up-conversion, so that the wavelengths remain unchanged after each step of the PDU. The DPUC resolves the problem of the increasing photon wavelengths as the DPDC processes cascade, and the long wavelength exceeds the transparency window of materials eventually and raises challenges for single-photon detector technology. With the photon wavelength unchanged, we can use a standard design as a fundamental building block, for all PDUs at different stages, which is important for scalable photonic integration.

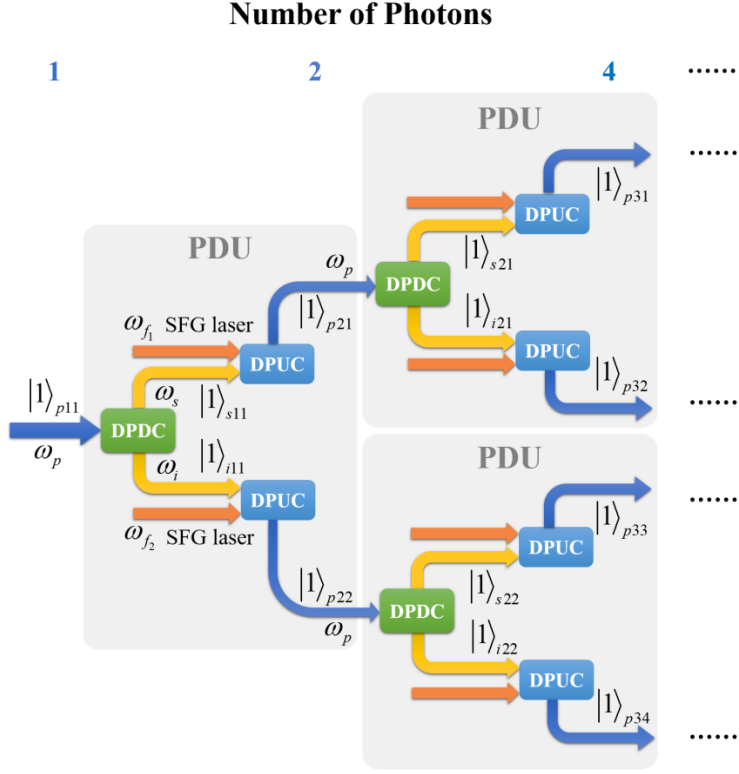

**Fig. 1** Scheme for deterministic *N*-photon state generation using PDU. As marked in a grey box, a PDU is used to convert a pump photon to biphoton with the same frequency deterministically, through nondegenerate DPDC and DPUC processes. By cascading the PDUs, the photon number can be doubled exponentially, towards deterministic generation of an *N*-photon state, with the Fock state as an example.

The practical layout of each PDU can be illustrated in Fig. 2(a), in the form of LNOI circuits. We choose a quasi-phase-matched (QPM) high-$Q$ microring resonator for DPDC process. The reason is in two folds. Firstly, the microring resonator can greatly enhance the photon interaction in a small footprint, and our modeling and calculation shows DPDC can be achieved, considering the ultra-high nonlinearity in LNOI. Secondly, the cavity enhancement can keep the single photon spectra unbroadened with proper resonator design, while in the case of non-resonant parametric down-conversion process, the photon bandwidth is determined only by the phase-matching bandwidth, which is normally on the order of hundreds of GHz.[31, 32] It is a fundamental challenge to manage the phase matching over such increasingly broadened photon bandwidth.

In our layout, the DPDC is nondegenerate in frequency, so that the signal and idler photons can be separated by the on-chip wavelength division multiplexing (WDM1) device. Such WDM devices have been reported in LNOI, using different designs including Mach-Zehnder interferometer,[33] multimode interferometer,[34] etc. Then the parametric photons enter centimeter-long periodically poled LNOI (PPLNOI) spiral waveguides for DPUC. With LNOI waveguide of such length, our calculation shows that unitary up-conversion efficiency can be achieved via a sum-frequency generation (SFG) process with a single-mode SFG laser and only mW-level continue-wave (cw) power. Different SFG laser wavelengths are chosen for the signal and idler photons, so that they are up-converted to the same wavelength as the pump, before entering the PDUs in the next stage. The residue SFG laser is rejected using WDM2/WDM3 from the up-converted single/idler photons, respectively, and may be reused to up-convert the single/idler photons PDUs in the next stage. Only two SFG laser wavelengths are required over the whole *N*-photon generation chip, that greatly simplify the setup. An *N*-photon state generation can be achieved by the integration of these PDUs, and the example in Fig. 2(b) shows the Fock state generation by direct cascading PDUs.

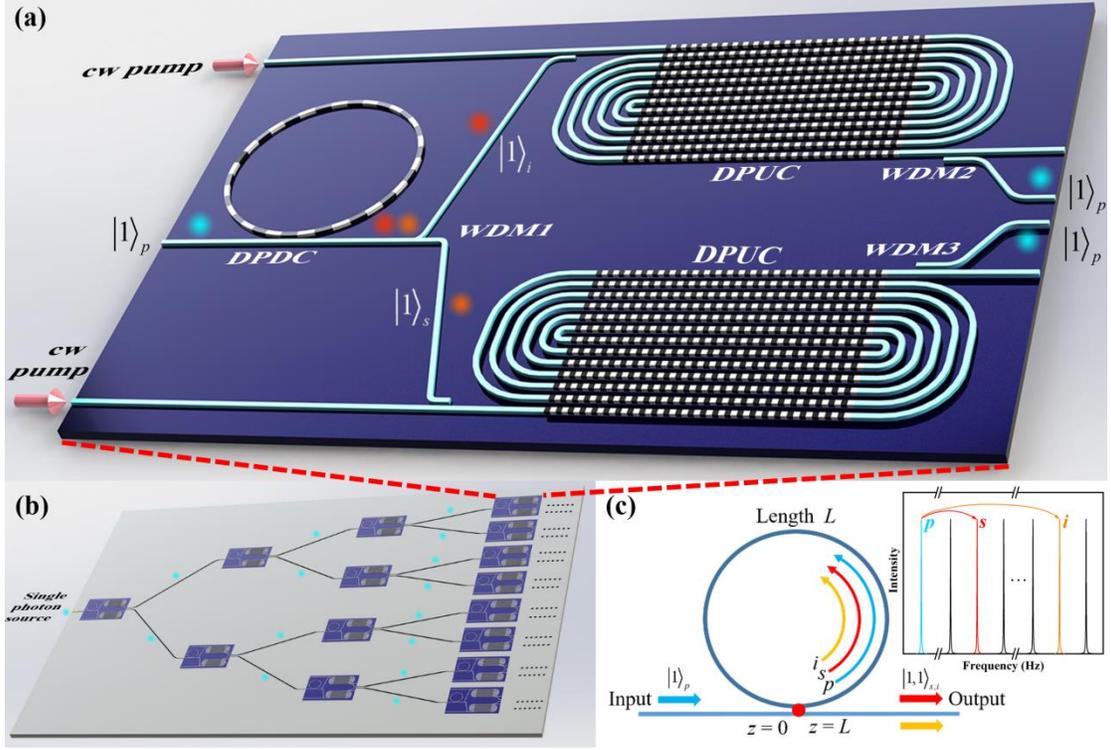

**Fig. 2** The example for PDU and *N*-photon state realization on a LNOI chip. (a) The PDU layout. The DPDC is implemented by a PPLNOI microring resonator and the DPUC is realized by PPLNOI spiral waveguides. (b) On-chip scheme for *N*-photon state generation using cascaded PDUs. (c) A simplified model of the DPDC process in a microring resonator. Here we model the DPDC as one-dimensional interaction following the propagation of pump, signal, and idler photons, where coupling point is marked in red for $z = 0$. The coordinate in the resonator $z$ varies between 0 and $L$, where $L$ is the resonator length. The inset shows longitudinal modes for pump, signal, and idler photons.

We model the DPDC process using the cavity-enhanced dual potential operators $\hat{\Lambda}(z,t)$ for quantization of the electromagnetic field,[35-37] and Hamiltonian can be written as the sum of linear ($\hat{H}_L$) and nonlinear ($\hat{H}_{NL}$) terms [35, 36] (See supplemental material A for details):

$$\hat{H} = \hat{H}_{NL} + \hat{H}_L$$
$$= \frac{A_{\text{eff}}}{3} \cdot \left( \frac{-\chi^{(2)}}{\varepsilon_0^2 n_p^2 n_s^2 n_i^2} \right) \int_L dz \frac{\partial \hat{\Lambda}_s}{\partial z} \frac{\partial \hat{\Lambda}_i}{\partial z} \frac{\partial \hat{\Lambda}_p}{\partial z} + \sum_{j=p,s,i} \frac{A_j}{2} \int_L dz \left[ \frac{1}{\varepsilon_0 n_j^2} \left( \frac{\partial \hat{\Lambda}_j}{\partial z} \right)^2 + \mu_0 \left( \frac{\partial \hat{\Lambda}_j}{\partial t} \right)^2 \right]. \quad (1)$$

It differs from the normal SPDC model with all the interacting light field quantized including the pump using $\hat{\Lambda}_j(z,t)$, where $j = p$, $s$, and $i$ represent pump, signal and idler, respectively. $\hat{\Lambda}(z,t)$ is the corresponding operator of the vector field $\Lambda$, which is defined by the electric displacement field $\mathbf{D}$, with $\mathbf{D} \equiv \nabla \times \Lambda$.[35-37] We use the "modes of the universe approach" to achieve $\hat{\Lambda}_j(z,t)$,[35] with

$$\hat{\Lambda}_j(z,t) = \int_0^\infty dk_j \sqrt{\frac{\hbar}{2\mu_0 \omega_j A_j}} \cdot \left( \alpha_{k_j} \hat{a}_{k_j} e^{ik_j z} e^{-i\omega_j t} + h.c. \right). \tag{2}$$

Here, $\alpha_{k_j} = i\sqrt{T_j} / \left[ \sqrt{2\pi} \left( 1 - \sqrt{1-T_j} e^{-ik_j L} \right) \right]$ represents for the cavity enhancement factor, with $T$ denotes the transmissivity of the resonator. $\hat{a}_{k_j}$ stands for the photon annihilation operator of wave vector $k_j$, $\omega_j$ is the angular frequency, and $h.c.$ represents for the Hermitian conjugate. $A_{\text{eff}} \equiv \iint_S dxdy U_p(x,y) U_s^*(x,y) U_i^*(x,y)$ is the effective spatial overlap, and $A_j \equiv \iint dxdy |U_j(x,y)|^2$ is the mode area for photon $j$, with $S$ denoting cross-section of the waveguide and $U(x, y)$ denoting the normalized transverse mode distribution of electrical field. And $L$, $n_j$ correspond to the length of the microring resonator and the effective refractive index of photon $j$, respectively.

To have unbroadened spectra for $N$-photon state generation, single-longitudinal-mode (SLM) oscillation must be achieved in this cavity-enhanced case, which requires the difference of the free spectrum range (FSR) of signal and idler light to be larger than linewidth of the cavity resonances.[38, 39] Such condition is easily satisfied in the high-$Q$ case, where such high $Q$-factor is also necessary for the high efficiency as required by DPDC on the other hand. At certain resonator size and dispersion, the $Q$ requirement of SLM oscillation can be derived as (See details in supplemental materials section B):

$$Q > \frac{\omega R}{c\left|1/n_{s0} - 1/n_{i0}\right|},\qquad(3)$$

where $R$ is the radius of the microring, with $L = 2\pi R$. As expected, smaller resonator size and larger dispersion can relax the requirement of $Q$ for SLM oscillation. Assuming a cross-section structure with 60° wall slope following the normal LNOI fabrication technique [28, 40, 41] (inset of Fig. 3(b), and such cross-section is used for all the following simulation) and a mirroring radius of 30 μm, SLM oscillation can be achieved when $Q > 2.7 \times 10^4$, for the PDC process 646.91 nm → 1276.80 nm + 1311.29 nm.

Under SLM oscillation condition, it is reasonable to narrow the integration range for $k_j$ in Eq. (2) to around a single longitude mode $\Delta k_{FSR} \equiv 2\pi/L = 1/R$, which is chosen to be $2\Delta = 1/(\pi R)$ here. In addition, $k_j$, $n_j$ and $A_{eff}$ can be approximated as constants at center frequencies. Consequently, the dual potential can be rewritten as

$$\hat{\Lambda}_j(z,t) = \int_{k_{j0}-\Delta}^{k_{j0}+\Delta} dk_j \sqrt{\frac{\hbar}{2\mu_0 \omega_j A_j}} \cdot \left(\alpha_{k_j} \hat{a}_{k_j} e^{ik_j z} e^{-i\omega_j t} + h.c.\right).\qquad(4)$$

Thus the Hamiltonian becomes (See supplemental material section C for details)

$$\hat{H} = g \int_{k_{p0}-\Delta}^{k_{p0}+\Delta} dk_p \int_{k_{s0}-\Delta}^{k_{s0}+\Delta} dk_s \int_{k_{i0}-\Delta}^{k_{i0}+\Delta} dk_i \cdot \left(v(k_p) v^*(k_s) v^*(k_i) \hat{a}_{k_s}^+ \hat{a}_{k_i}^+ \hat{a}_{k_p} e^{-i(\omega_p - \omega_s - \omega_i)t} + h.c.\right)$$
$$+ \sum_{j=p,s,i} \int_{k_{j0}-\Delta}^{k_{j0}+\Delta} dk_j \frac{\hbar c k_j R}{n_j} \left|v(k_j)\right|^2 \hat{a}_{k_j}^+ \hat{a}_{k_j} \qquad(5)$$

where $g = \left(\chi^{(2)} \sqrt{\hbar/\varepsilon_0 \pi}\right) \cdot \left(c/n_{p_0} n_{s_0} n_{i_0}\right)^{3/2} \cdot \left(A_{eff}/\sqrt{A_{p_0} A_{s_0} A_{i_0}}\right) \sqrt{k_{p_0} k_{s_0} k_{i_0}} \, R/12$ and $v(k_j) \equiv \sqrt{T_j}/\left(1 - \sqrt{1-T_j} e^{-ik_j \cdot 2\pi R}\right)$.

To further simplify the Hamiltonian, we introduce the "normalized discrete Hilbert-space photon annihilation operator" $\hat{\varsigma}_j = \sqrt{R} \int_{k_{j0}-\Delta}^{k_{j0}+\Delta} dk_j v(k_j) \hat{a}_{k_j}$ [35] and get (See supplemental material section C for details):

$$\frac{\hat{H}}{\hbar} = \xi\left(\hat{\varsigma}_s^+\hat{\varsigma}_i^+\hat{\varsigma}_p + \hat{\varsigma}_p^+\hat{\varsigma}_s\hat{\varsigma}_i\right) + \omega_{p_0}\hat{\varsigma}_p^+\hat{\varsigma}_p + \omega_{s_0}\hat{\varsigma}_s^+\hat{\varsigma}_s + \omega_{i_0}\hat{\varsigma}_i^+\hat{\varsigma}_i, \tag{6}$$

where $\xi$ is the nonlinear interaction strength:

$$\xi \equiv \frac{\chi^{(2)}}{12}\sqrt{\frac{\hbar}{\pi\varepsilon_0 R}}\left(\frac{c}{n_{p_0}n_{s_0}n_{i_0}}\right)^{\frac{3}{2}}\frac{A_{\text{eff}}}{\sqrt{A_{p_0}A_{s_0}A_{i_0}}}\sqrt{k_{p_0}k_{s_0}k_{i_0}}. \tag{7}$$

Then we use stationary Schrödinger equation [37, 42] to calculate the time evolution of quantum states inside the microring. With single-photon pump input $|1\rangle_p = \hat{\varsigma}_p^+|0\rangle$, the time-independent Hamiltonian in Eq. (6) can be derived (See supplemental materials section D for detail)

$$|\Psi(t_I)\rangle = e^{-i\omega_{p_0}t_I}\left[\cos(\xi t_I)|1\rangle_p - i\sin(\xi t_I)|1,1\rangle_{s,i}\right], \tag{8}$$

where $|1,1\rangle_{s,i} = \hat{\varsigma}_s^+\hat{\varsigma}_i^+|0\rangle$ represents the photon pairs generation, and $t_I$ is the nonlinear interaction time, which can be approximated by $t_I \approx Ft_r$, where $F$ is the finesse of the resonator and $t_r$ is the roundtrip time of the resonator.[38] This nonlinear interaction time $t_I$ is equal to $2\pi$ times the average photon intercavity lifetime. The efficiency for the single photon PDC is thus

$$\eta_{\text{PDC}} = \sin^2\left(\frac{2\pi\xi Q}{\omega}\right). \tag{9}$$

Therefore, unitary conversion efficiency can be achieved when $Q$ is high enough.

Taking the 30-μm radius LNOI microring resonator as an example, we calculate the required $Q$ value for DPDC (See supplemental materials section E for more detail). Following Eq. (7), we first obtain $\xi$ by calculating refractive indices $n_{j0}$, wave vectors $k_{j0}$, effective spatial overlap $A_{\text{eff}}$, and mode area $A_j$ for different wavelength. Substituting $\xi$ into Eq. (9), we obtain the relation between $Q$ and $\eta_{\text{PDC}}$ under different wavelength, as shown in Fig. 3(a). For example, for the PDC process 646.91 nm → 1276.80 nm + 1311.29 nm, DPDC can be achieved with relatively low $Q$ of $4.11\times10^7$ at 1311.29 nm, which is experimentally feasible considering the $Q$ over $10^8$ for LNOI resonator in experiment realizations.[43-45] We also calculate the required $Q$ for DPDC under different microring-radius $R$, which shows that the $Q$-requirement can be relaxed by smaller $R$, as shown in Fig. 3(b).

We also model the DPUC process as the single photon sum-frequency (SFG) generation with classical laser light. Here, we take the DPUC process with SFG1 laser as example, and the other DPUC process can be calculated using the same method. The SFG1 laser light can be considered as non-dissipative, so that the Hamiltonian of the up-conversion process can be written as:[46, 47]

$$\hat{H}_\mathrm{I} = i\hbar \iota \left( \hat{a}_s \hat{a}_p^+ - h.c. \right), \tag{10}$$

where $\iota = \sqrt{\kappa P_\mathrm{SFG1}}$ is the effective SFG nonlinear strength with $P_\mathrm{SFG1}$ represents the power of the SFG1 laser, and $\kappa \equiv 2\varepsilon_0^{1/2} \mu_0^{3/2} \left( \omega_s \omega_p / n_s n_\mathrm{SFG1} n_p \right) d_\mathrm{eff}^2 \left( A_\mathrm{eff} / \sqrt{A_s A_\mathrm{SFG1} A_p} \right)^2$, where $d_\mathrm{eff}$ represents the effective nonlinear coefficient. In our case, the integration area for $A_\mathrm{eff}$ is the cross-section of the LNOI waveguide. Basing on Eq. (10), for an input low frequency photon, the up-conversion efficiency can be written as:[46, 48]

$$\eta_\mathrm{PUC} = \sin^2 \left( \sqrt{\kappa P_\mathrm{SFG1}} l \right), \tag{11}$$

where $l$ is the waveguide length.

Then, we calculate parametric up-conversion efficiency under different SFG laser power and waveguide length (See detail calculation in supplemental material E). To consist with the DPDC process, we choose the 1276.80 nm (single photon) + 1311.29 nm (SFG1 laser) → 646.91 nm and the 1276.80 nm (SFG2 laser) + 1311.29 nm (single photon) → 646.91 nm process. The calculation results are shown in Fig. 3(c) and (d), where the blue curve represents the condition for DPUC.

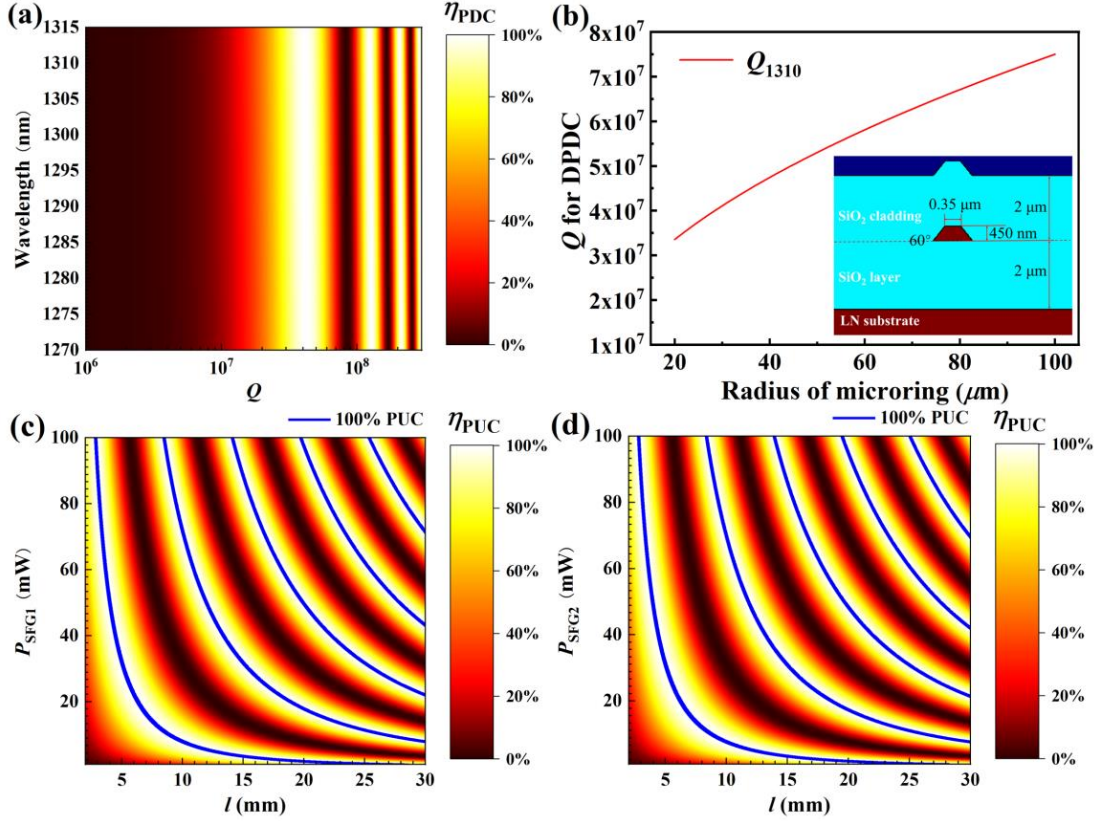

**Fig. 3** Calculation results of the PDC and PUC efficiency in the LNOI circuit. (a) The relation between PDC efficiency $\eta_{PDC}$ and $Q$ for parametric light, the ripple of the conversion efficiency arises from the Rabi-like oscillation between $|1\rangle_p$ and $|1,1\rangle_{s,i}$. (b) The relation between the required $Q$ for DPDC and the radius of the microring. The inset is the transverse structure of the simulated LNOI waveguide. (c) The up-conversion efficiency $\eta_{PUC}$ with different waveguide length $l$ and SFG1 laser power $P_{SFG1}$. The blue curve indicates DPUC, where multiple DPUC lines indicates Rabi-like oscillation between low frequency and high frequency photons. (d) The up-conversion efficiency $\eta_{PUC}$ with different waveguide length $l$ and SFG2 laser power $P_{SFG2}$.

The DPUC has already been achieved in many platforms, like reverse-proton-exchanged (RPE) [48-50] and Ti-diffused [51] PPLN waveguides with hundreds milliwatts of SFG laser power and several centimeters of waveguide, while for LNOI circuit, $P_{SFG1} \approx 8$ mW with $l = 1$ cm is sufficient according to our calculation. Such power can be achieved by laser diode directly, hence exempting the need for Erbium-doped fiber amplifier (EDFA). Taking advantage of the strong refractive index difference of LNOI waveguide, cm-order waveguide can be integrated in a small area on-chip through small-bending-radius spiral structure. For example, for the waveguide with geometry as shown in the inset of Fig. 3(b), 1-cm length waveguide can be achieved within 1 mm × 1 mm area using 100 μm-bending-radius spiral waveguide,[52] while the bending loss is still negligible. Combining the DPUC spiral waveguide with the DPDC microring resonator, PDU can be achieved on chip with extremely small footprint.

Besides the *N*-photon Fock state, many other *N*-qubit states, which are the key resource for practical quantum technology applications, can also be generated deterministically using PDUs. As examples shown in Fig. 4, we propose the on-chip design for *N*-photon cluster states and GHZ states, which are the key for one-way quantum computation,[53-55] and useful for quantum communication,[3, 4, 20] respectively. Here, we code these states on path, because it is a highly scalable and easily manipulable degree of freedom in circuits. In the design, the increase of photon number is achieved by PDUs, while state encoding is realized by optical interference and phase control through on-chip beam splitters and phase modulators using devices like multi-mode interference (MMI) couplers [56] and electro-optic modulation,[41, 57, 58] respectively. To achieve interference between paths that are not next to each other, crossers based on MMI or tapers etc.[41, 59] are used. After two stages of PDUs, the 4-photon cluster state and 4-photon GHZ state are generated as shown in Fig. 4(a) and Fig. 4(b), respectively (see supplemental materials F and G for details):

$$|\text{cluster}_4\rangle = \frac{1}{2}\left(|\tilde{0}\tilde{0}\tilde{0}\tilde{0}\rangle_{1234} + |\tilde{0}\tilde{0}\tilde{1}\tilde{1}\rangle_{1234} + |\tilde{1}\tilde{1}\tilde{0}\tilde{0}\rangle_{1234} - |\tilde{1}\tilde{1}\tilde{1}\tilde{1}\rangle_{1234}\right), \quad (12)$$

$$|\text{GHZ}_4\rangle = \left(|\tilde{0}\rangle^{\otimes 4} + e^{i\varphi}|\tilde{1}\rangle^{\otimes 4}\right)/\sqrt{2}, \quad (13)$$

where $|\tilde{0}\rangle_j$ and $|\tilde{1}\rangle_j$ refer to the *j*th qubit defined from the corresponding two different paths. Scaling the basic unit of 4-photon cluster state up with proper linear optical operation, or cascading more stages of PDUs following the 4-photon GHZ state design, $N$-photon cluster states or GHZ state $|\text{GHZ}_N\rangle = \left(|\tilde{0}\rangle^{\otimes N} + e^{i\varphi}|\tilde{1}\rangle^{\otimes N}\right)/\sqrt{2}$ can be obtained respectively. Generally, PDUs together with proper linear optical devices is a universal scheme for deterministic generation of arbitrary $N$-qubit states, where up to $2^M$-photon states can be generated using $M$ stages of PDUs.

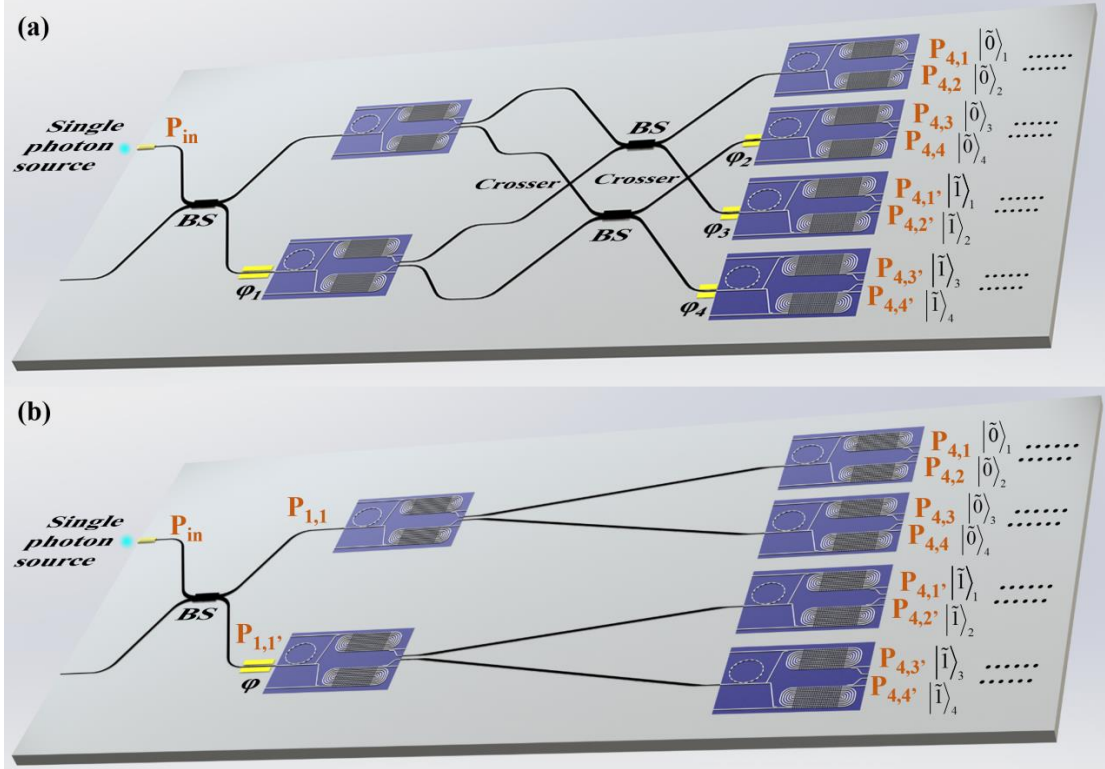

**Fig. 4** Circuit design for generating different $N$-qubit states. (a) Circuit design for $N$-photon cluster state. BSs are 50:50 beam splitters for photon separation and interference, and crossers are used for two waveguides to intersect with negligible crosstalk. After 2 stages of PDU, a 4-photon cluster state can be generated, in which $\varphi_1 = \pi/2$, $\varphi_2 = 7\pi/4$, $\varphi_3 = \pi/2$ and $\varphi_4 = \pi/4$, (b) Circuit design for $N$-photon GHZ state, where after 2 stages, a 4-photon GHZ state can be generated. The darkened part in the figure is the PDU, of which the detailed structure is shown in Fig. 2(a).

In conclusion, we present a scheme for arbitrary $N$-photon state generation with unlimited photon number in principle, where $N$-photon Fock state, GHZ state and cluster state are taken as examples to demonstrate the detail design. Such scheme, utilizing the high-nonlinearity LNOI circuit, makes large-size quantum state generation experimentally feasible for the first time. The key component in the design is an ensemble of scalable standard basic units called PDUs. Basing on our calculation, in the unit, DPDC and DPUC can be achieved with ~$4\times10^7$ $Q$-factor microring resonators, 1-cm-long waveguides and 8-mW SFG power in this unit, respectively. These numbers have been reported separately in the existing experimental papers on LNOI.[28, 60, 61] The strong field confinement of LNOI also enables small PDU footprint for the potential large scale integration, paving the route to the large photon number generation on a single chip. The remaining challenges for the experimental demonstration are technical problems, which are not unrealistic in principle, including resonance matching and fabrication error control in the PDUs, etc. We show the strong single-photon interaction is possible in LNOI devices, and it is utilized for the *N*-photon states generation but may not be limited for this application as discussed here. This strong single-photon interaction can also be used for photon manipulation to realize quantum gates, quantum storage and so on, to push forward the development of quantum computation,[21] quantum communication [4, 20] and the overall quantum information technology.

*Disclosures*

The authors declare no conflicts of interest.

*Acknowledgments*

This work was supported by the National Key R&D Program of China (No. 2019YFA0705000), Key R&D Program of Guangdong Province (No. 2018B030329001), Leading-edge technology Program of Jiangsu Natural Science Foundation (No. BK20192001), National Natural Science Foundation of China (51890861, 11690033, 11974178), the Excellent Research Program of Nanjing University (ZYJH002). H. Y. L. gratefully acknowledges the support of the National Postdoctoral Program for Innovative Talents (BX2021122).

*Code, Data, and Materials Availability*

Data underlying the results presented in this paper are not publicly available at this time but may be obtained from the authors upon reasonable request.

**Caption List**

**Fig. 1** Scheme for deterministic *N*-photon state generation using PDU. As marked in a grey box, a PDU is used to convert a pump photon to biphoton with the same frequency deterministically, through nondegenerate DPDC and DPUC processes. By cascading the PDUs, the photon number can be doubled exponentially, towards deterministic generation of an *N*-photon state, with the Fock state as an example.

**Fig. 2** The example for PDU and *N*-photon state realization on a LNOI chip. (a) The PDU layout. The DPDC is implemented by a PPLNOI microring resonator and the DPUC is realized by PPLNOI spiral waveguides. (b) On-chip scheme for *N*-photon state generation using cascaded PDUs. (c) A simplified model of the DPDC process in a microring resonator. Here we model the DPDC as one-dimensional interaction following the propagation of pump, signal, and idler photons, where coupling point is marked in red for $z = 0$. The coordinate in the resonator $z$ varies between 0 and $L$, where $L$ is the resonator length. The inset shows longitudinal modes for pump, signal, and idler photons.

**Fig. 3** Calculation results of the PDC and PUC efficiency in the LNOI circuit. (a) The relation between PDC efficiency $\eta_{PDC}$ and $Q$ for parametric light, the ripple of the conversion efficiency arises from the Rabi-like oscillation between $|1\rangle_p$ and $|1,1\rangle_{s,i}$. (b) The relation between the required $Q$ for DPDC and the radius of the microring. The inset is the transverse structure of the simulated LNOI waveguide. (c) The up-conversion efficiency $\eta_{PUC}$ with different waveguide length *l* and SFG1 laser power $P_{SFG1}$. The blue curve indicates DPUC, where multiple DPUC lines indicates Rabi-like oscillation between low frequency and high frequency photons. (d) The up-conversion efficiency $\eta_{PUC}$ with different waveguide length *l* and SFG2 laser power $P_{SFG2}$.

**Fig. 4** Circuit design for generating different *N*-qubit states. (a) Circuit design for *N*-photon cluster state. BSs are 50:50 beam splitters for photon separation and interference, and crossers are used for two waveguides to intersect with negligible crosstalk. After 2 stages of PDU, a 4-photon cluster state can be generated, in which $\varphi_1 = \pi/2$, $\varphi_2 = 7\pi/4$, $\varphi_3 = \pi/2$ and $\varphi_4 = \pi/4$, (b) Circuit design for *N*-photon GHZ state, where after 2 stages, a 4-photon GHZ state can be generated. The darkened part in the figure is the PDU, of which the detailed structure is shown in Fig. 2(a).